\documentclass[11pt]{article}
\usepackage{moriond,epsfig}
\usepackage{fancyheadings}

\def\be{\begin{equation}}
\def\ee{\end{equation}}
\def\bea{\begin{eqnarray}}
\def\eea{\end{eqnarray}}

\newcommand{\pacz}{Paczy\'nski }
\newcommand{\macho}{\textsc{macho}}
\newcommand{\sg}{\mbox{sloan g'}}
\newcommand{\sr}{\mbox{sloan r'}}
\newcommand{\si}{\mbox{sloan i'}}

\begin{document}

\thispagestyle{fancy}
\chead[\fancyplain{}{\bfseries\thepage}]
{\begin{minipage}{1.9\textwidth}
\begin{tabular}{p{16cm}}
\\
\hline\\
\end{tabular}
\\
{\bf  To appear in the proceedings of the XXXIXth Rencontres de Moriond}\\
{\bf  ``Exploring the Universe", La Thuile, Italy, March 28-April 4, 2004}\\
\end{minipage}}

\vspace*{4cm}
\title{ THE POINT-AGAPE MICROLENSING SURVEY: FIRST CONSTRAINT ON MACHOS
  TOWARDS M31 }

\author{
St\'ephane PAULIN-HENRIKSSON, Sebastiano CALCHI NOVATI\\
on behalf of the POINT-AGAPE collaboration
}

\address{
PCC-Coll\`ege de France, 11 Place Marcelin Berthelot, 75231
  Paris cedex 5, France. \textsf{paulin@cdf.in2p3.fr}\\
ITP - Univ. of Zurich, Winterthurerstrasse 190 CH-8057 Zurich,
Switzerland. \textsf{novati@physik.unizh.ch}
}

\maketitle\abstracts{
To reveal the galactic dark matter in the form of \macho s ("Massive
Astrophysical Compact Halo Objects"), the POINT-AGAPE collaboration is carrying out a
search for gravitational microlensing towards M31. A clear
microlensing signal is detected. The high-threshold analysis of 3-year data leads to 7 bright and short
microlensing candidates. The preliminary estimation of the detection
efficiency implies that less
than 25\% (60\%) of standard halos can be composed of objects with a mass
between $10^{-4}\,M_\odot$ and $10^{-1}\,M_\odot$ ($10^{-1}\,M_\odot$ and
$1\,M_\odot$) at the 95\% C.L. This result is compatible with previous
microlensing results towards the Magellanic Clouds and gives the first
constraints on \macho s for a distant
spiral galaxy.
}

\section{Introduction}

\subsection{The galactic dark matter}
According to the first generation of microlensing experiments towards
the Magellanic Clouds, in particular EROS \cite{afonso03} and MACHO \cite{alcock00},
it is now clear that the halo of the Milky Way is not full of \macho s (``Massive Astrophysical Compact Halo
Objects''). However, galactic halos could still be partly composed of \macho
s (brown dwarfs,
Jupiter-like objects,\ldots). Indeed, if the missing baryons belong to galactic halos, they could
contribute typically up to 15\% to the halo masses.

\subsection{Gravitational microlensing towards M31} 
Microlensing surveys towards M31 probe the halos of M31
and the Milky Way. POINT-AGAPE and several other experiments are devoted to this line of
sight\cite{riffeser03,dejong04}. Although M31 is the nearest spiral galaxy, it is
 about 15 times further away than the Magellanic Clouds. The stars are not
 resolved and it is necessary to develop new analysis methods. As
 suggested by \mbox{P. Baillon} \mbox{et al. (1993) \cite{baillon93}}, % (see also \mbox{A. Gould (1996)} \cite{gould96}),
the
 AGAPE experiment implements the ``pixel lensing'' (i.e. the study
 of the CCD pixel lightcurves). As in most cases we do not have access to the
 magnitude of the sources, the \pacz{} curve \cite{paczynski86}
 (i.e. the lightcurve of a microlensing event with a pointlike source
 and lens) has only 3 clearly observable
 parameters in each filter: the date of maximum magnification ($t_0$), the timescale given by the Full-Width-At-Half-Maximum ($t_{1/2}$),
 and the flux increase ($\Delta\Phi$). Unless the
 lightcurve is very well sampled with a good signal-to-noise ratio, this implies a
 degeneracy between the Einstein crossing time, the impact
 parameter and the flux of the lensed star. As the Einstein
 crossing-time contains the physical information about the lens,
 the estimation of the optical depth can only be made with additional
 priors (e.g. the luminosity function of stars
 \cite{gondolo99}) and a good statistic.

\section{The POINT-AGAPE collaboration}

\subsection{Presentation}
The POINT-AGAPE collaboration carried out a photometric survey of M31
with the Wide Field Camera \mbox{(WFC)} on the $2.5\,$m Isaac Newton Telescope \mbox{(INT) \cite{webint}} at La
Palma, Canary Islands. It observed two fields of about
$0.3\,$deg$^2$ each, located North and South of the M31 center. The
pixel size is $0.33''$ and the survey is carried out in three filters:
\sg , \sr{} and \si , similar to the SDSS filters. The 3 first years
of data are analysed, consisting of 180 nights from
August 1999 to January 2002, with typically one hour per
night and seeing between $1''$ and $2''$. The analysis of
the fourth year is in progress.

\subsection{High threshold analysis of a three-year baseline}
\begin{figure}
\resizebox{\hsize}{!}{
\epsfig{figure=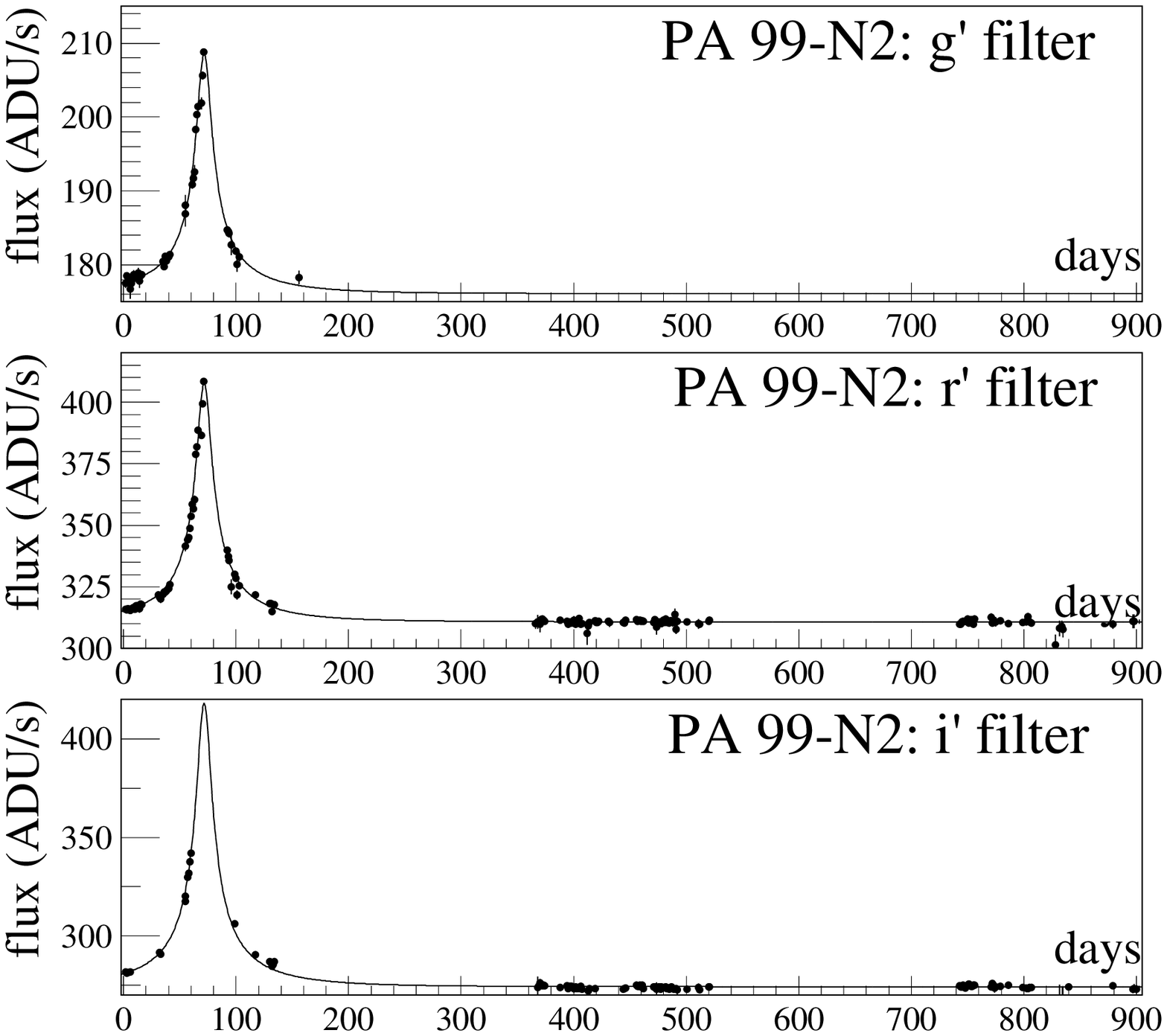}
\epsfig{figure=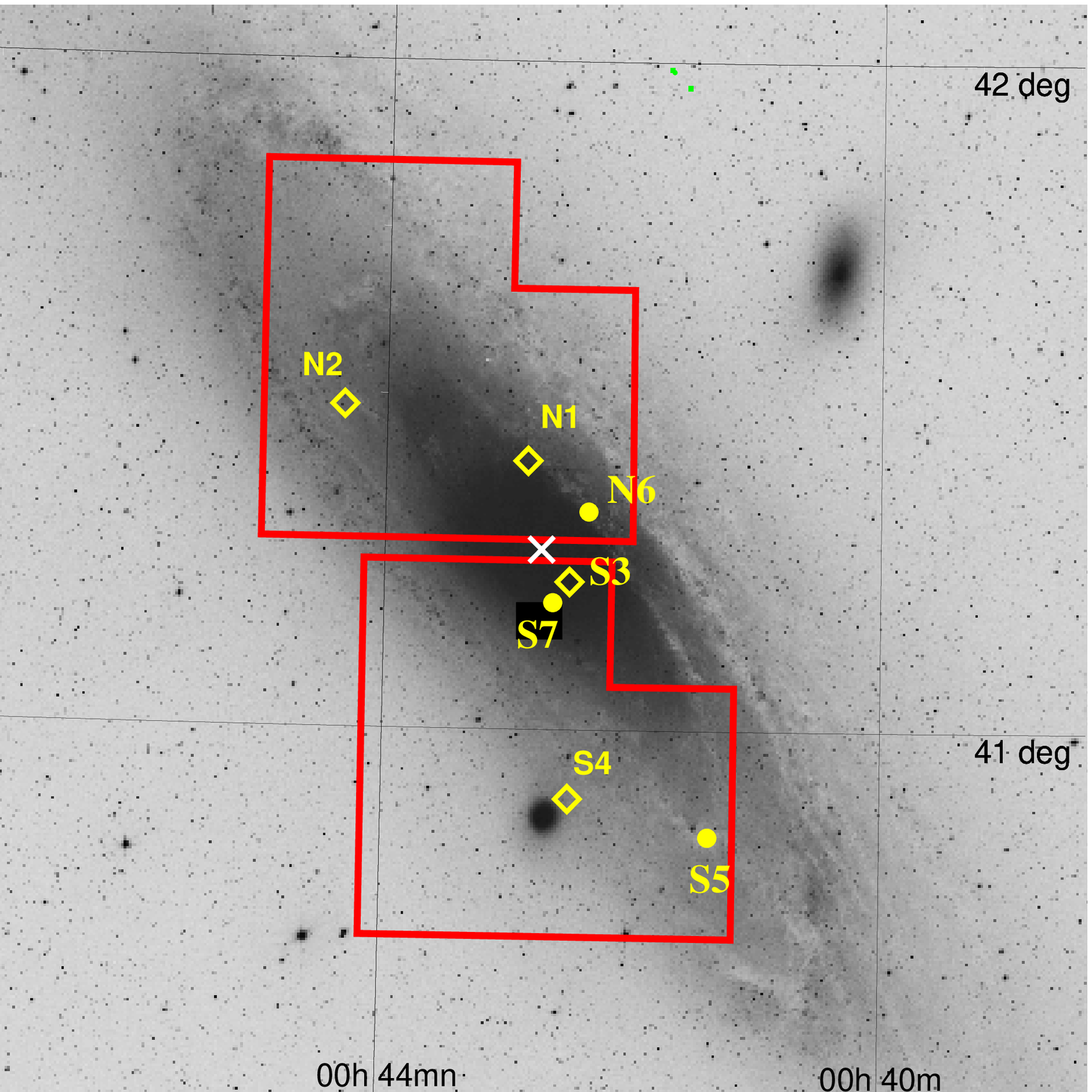}
}
\caption{
Left panel: the microlensing candidate PA$\,$99-N2
                                %\cite{paulin03,an03}
$^{13,3}$ in three filters (solid lines are the best-fit \pacz{}
curves %\cite{paczynski86}
$^{11}$). Right panel: positions of the 7 bright and short
microlensing candidates projected on M31 (yellow squarres and points
                                %\cite{paulin03,calchi04}
$^{13,7}$, red lines: boundaries of the observed field,
white cross: center of M31).}
\label{fig:n2-evts7-constraints}
\end{figure}

\begin{figure}
\resizebox{\hsize}{!}{
\epsfig{figure=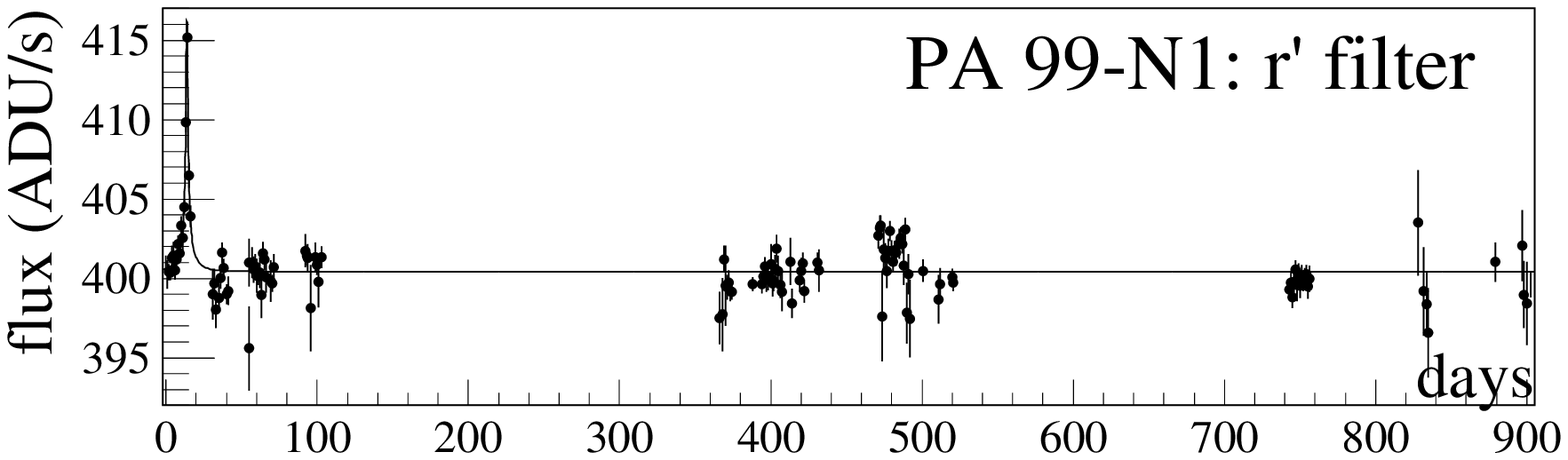}
\epsfig{figure=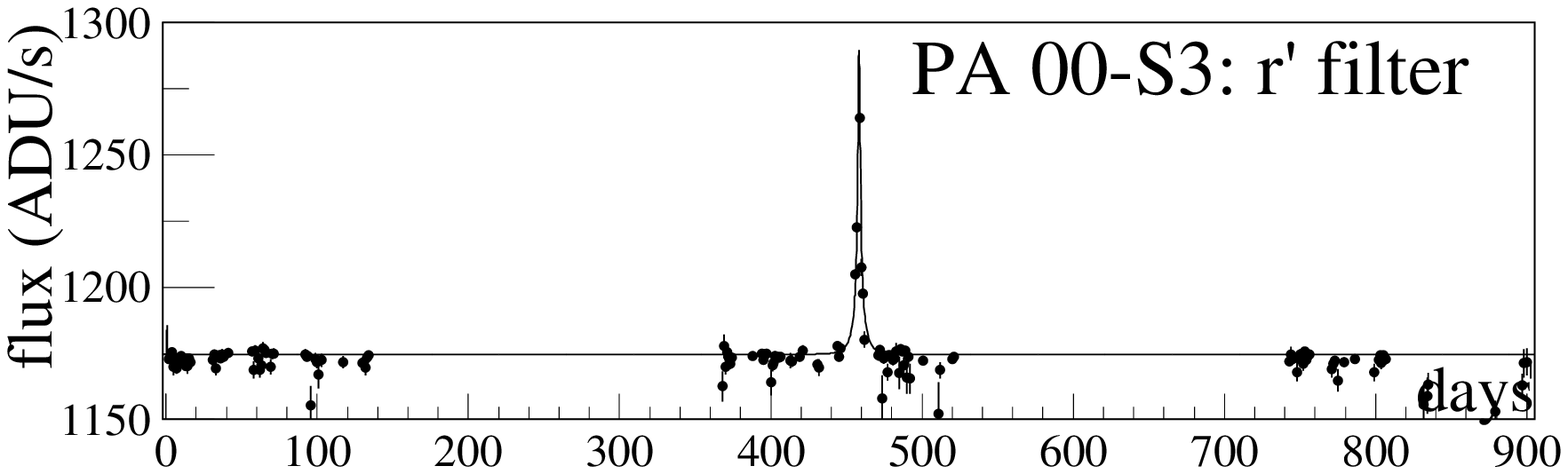}
}
\resizebox{\hsize}{!}{
\epsfig{figure=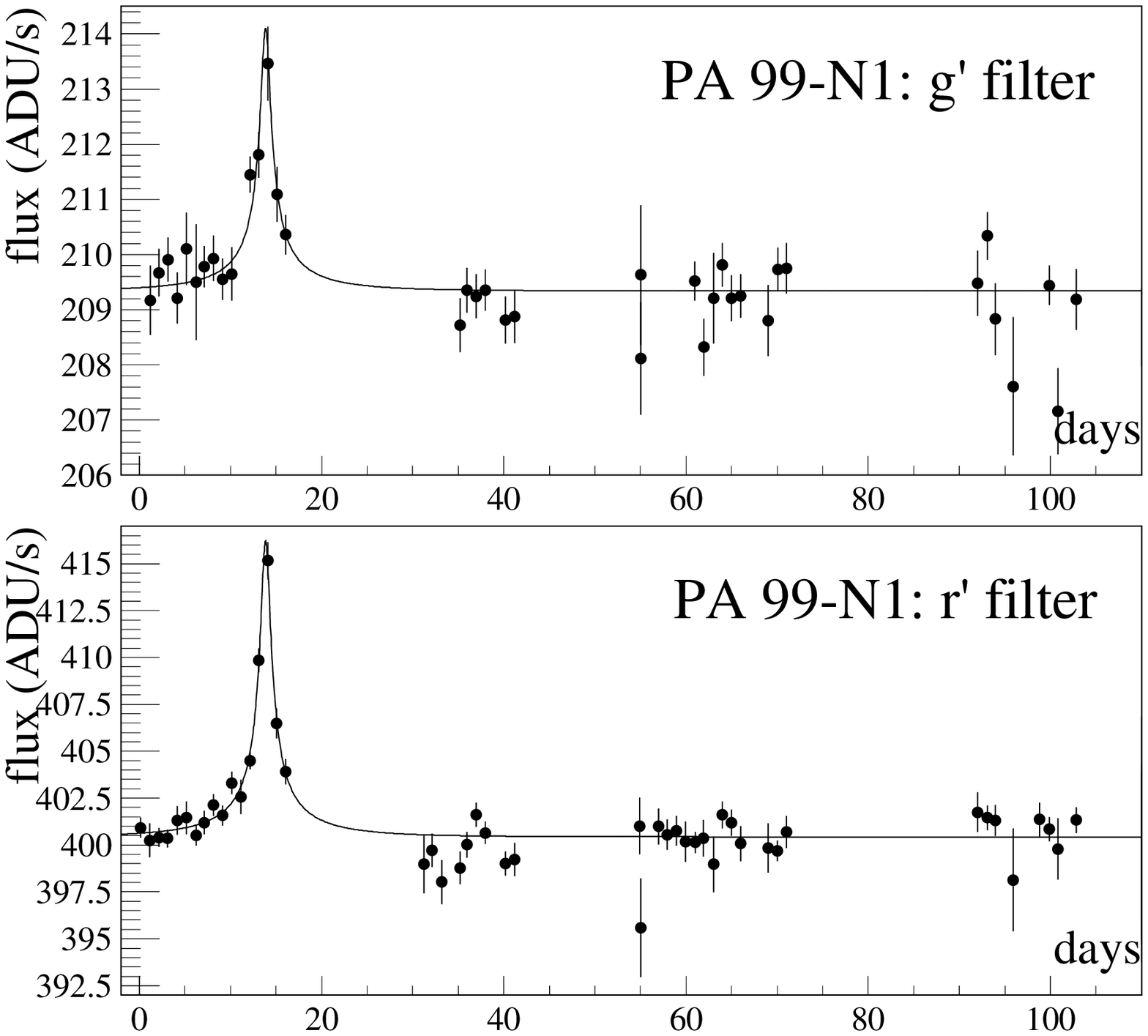}
\epsfig{figure=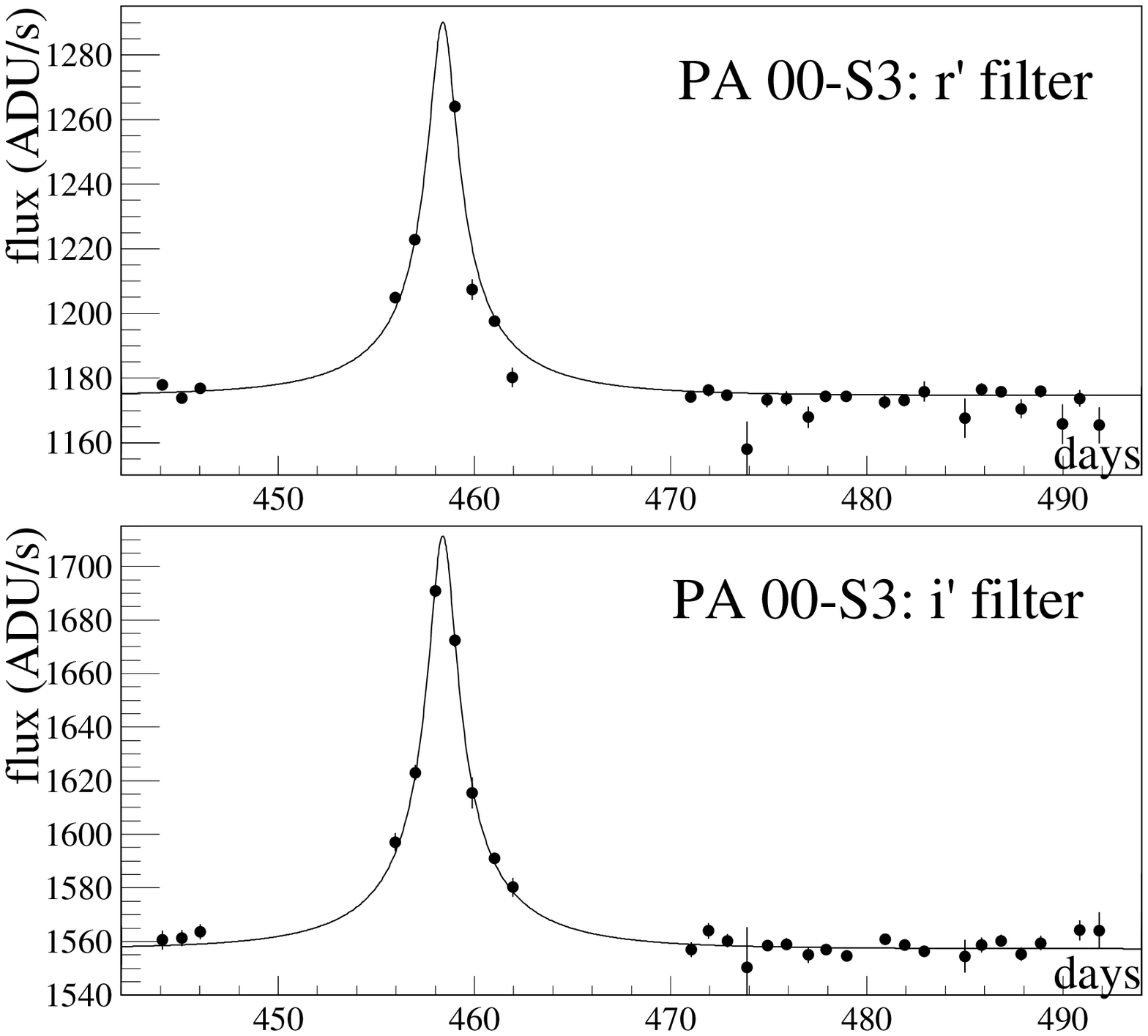}
}

\vspace*{1cm}

\resizebox{\hsize}{!}{
\epsfig{figure=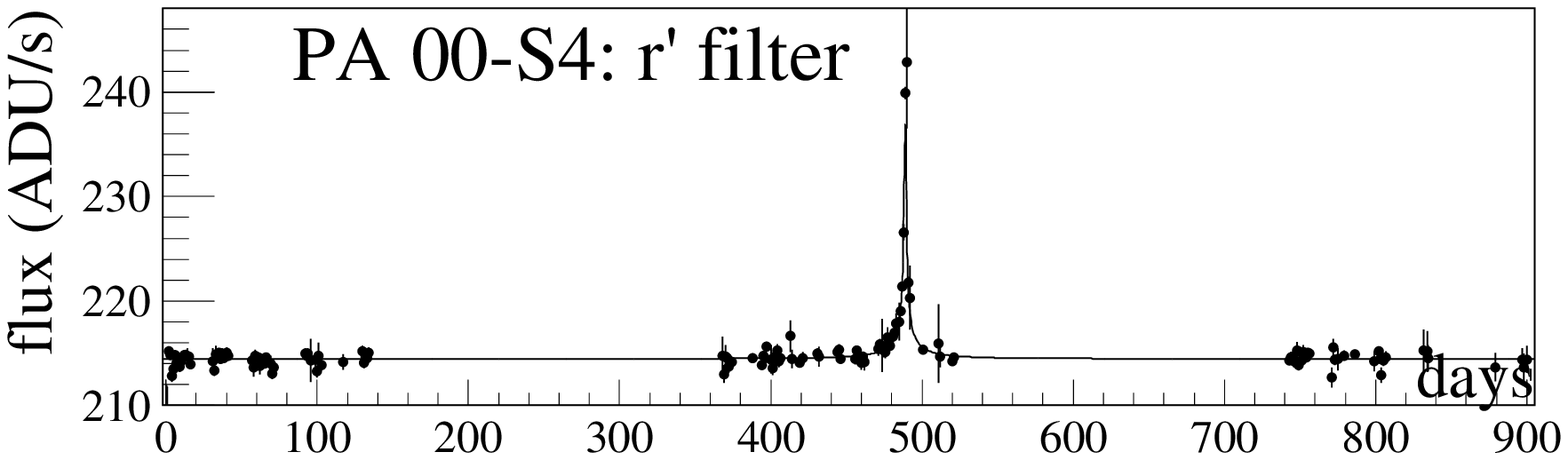}
\epsfig{figure=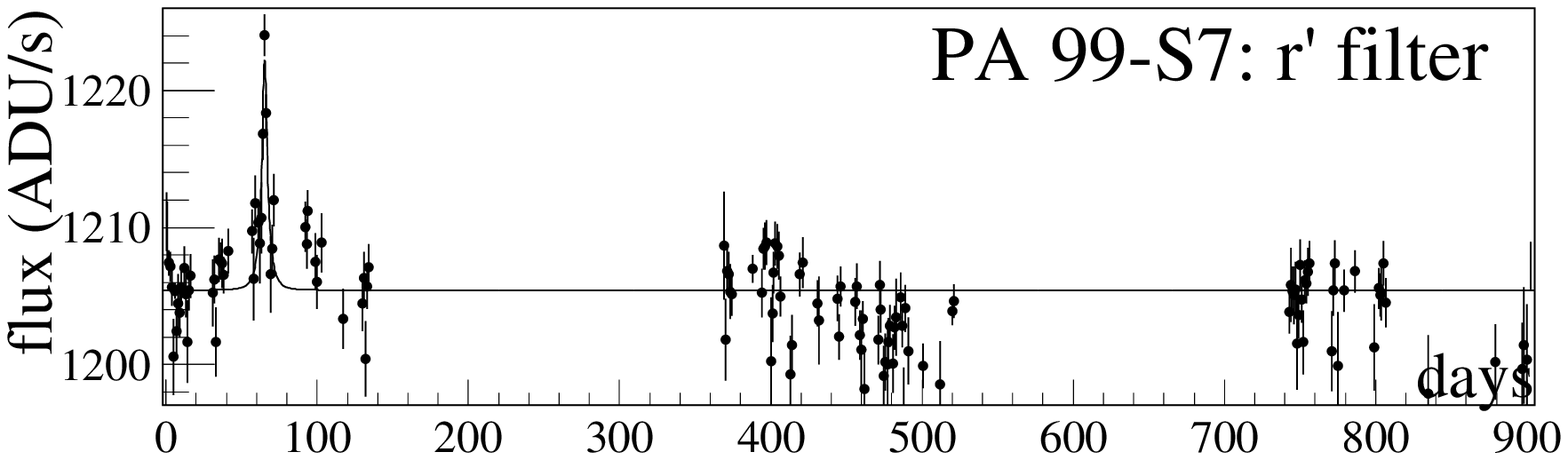}
}
\resizebox{\hsize}{!}{
\epsfig{figure=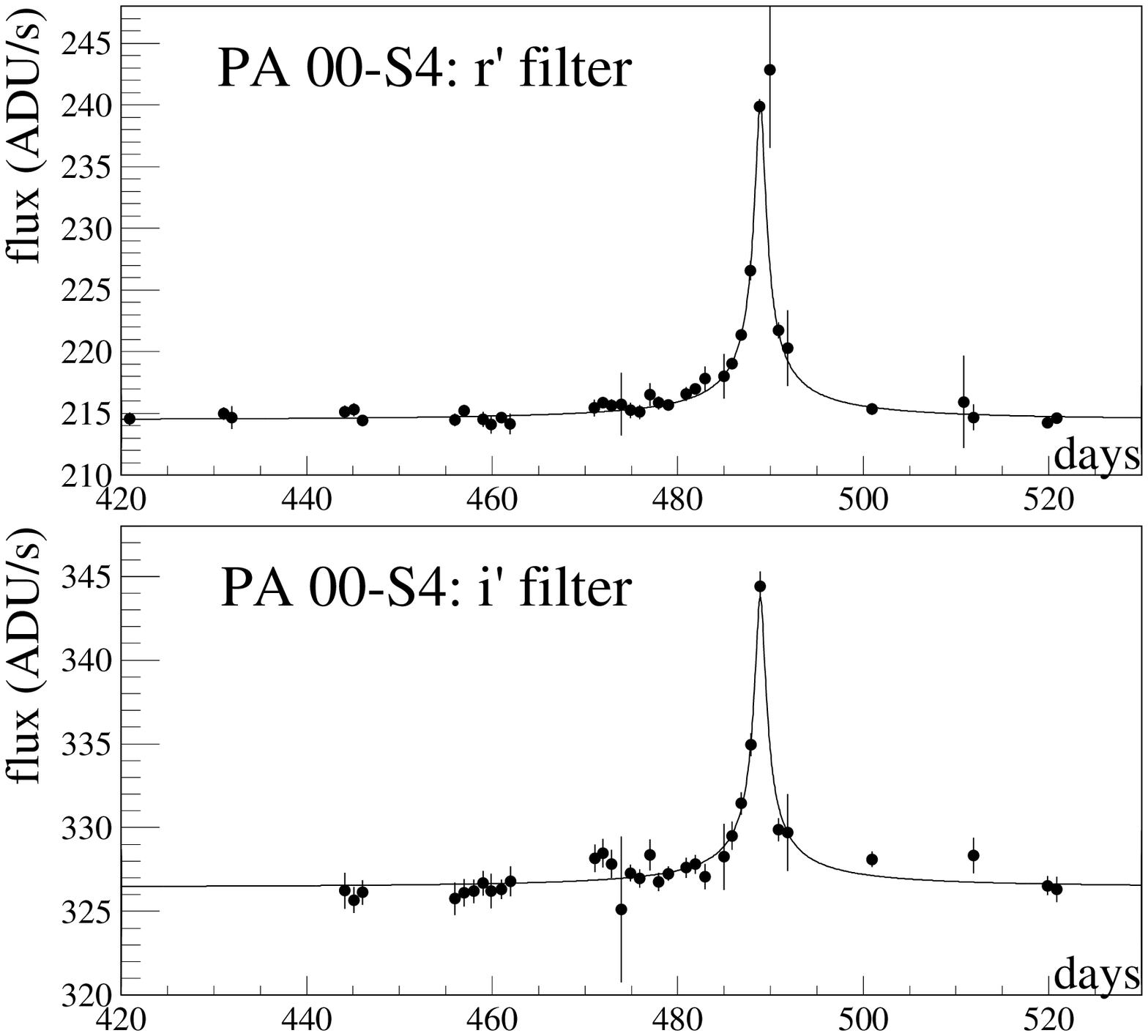}
\epsfig{figure=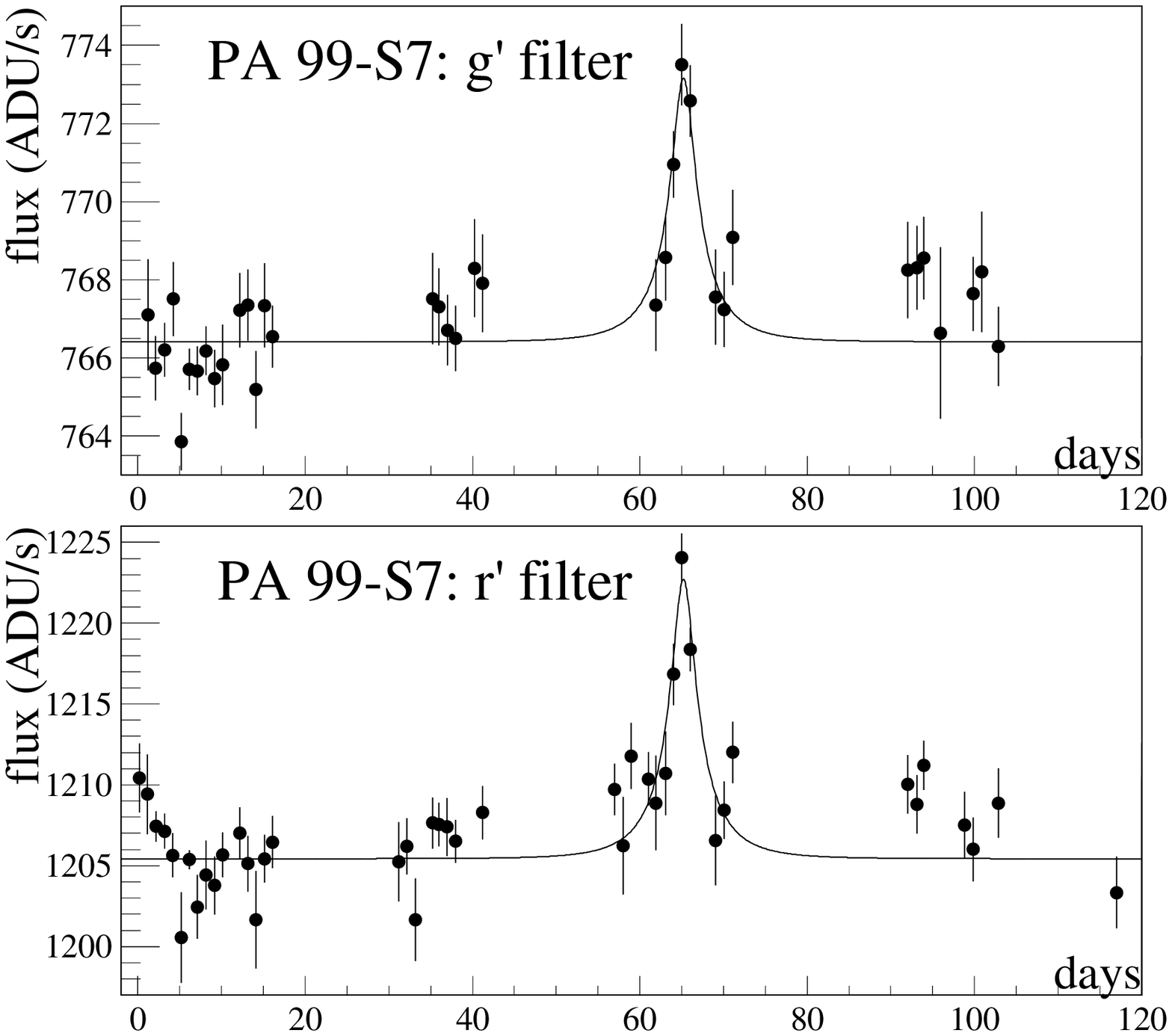}
}
\caption{Lightcurves of the four microlensing candidates:
  \mbox{PA$\,$99-N1 %\cite{paulin03,auriere01}
$^{13,5}$}, \mbox{PA$\,$00-S3 %\cite{paulin03}
$^{13}$}, \mbox{PA$\,$00-S4 %\cite{paulin03,paulin02}
$^{13,12}$} and \mbox{PA$\,$00-S7 %\cite{calchi04}
$^{7}$}. For each event, the top panel shows the 3-year baseline
  in the \mbox{$r'$} filter. Lower panels are zooms that focus on
  candidates in two bands. Solid lines are the best-fit \mbox{\pacz{}
    curves %\cite{paczynski86}
$^{11}$}.}
\label{fig:lightcurves}
\end{figure}

After the first steps of the data
reduction\cite{an04,paulin03}, we set up a catalogue of $\sim
5\times 10^4$ fluctuating lightcurves with $18.5<\Delta R<24$, %(full efficiency up to $\Delta R=22.5$),
$0<\Delta V-\Delta R<2$ and $-1<\Delta
R-\Delta I<2.5$ (where $\Delta V$, $\Delta R$ and $\Delta I$ are: $Z_p-2.5\log(\Delta\Phi)$ in the standard Jonhson-Cousins filters, and
$Z_p$ is the zero point). See
also Darnley et al. \mbox{(2004) \cite{darnley04}} for the specific
search for novae. This catalogue contains variable stars \cite{an04}
and contains a ten of convincing 
microlensing events. We search for them by fitting the lightcurves
to an achromatic \pacz{} curve \cite{paczynski86}, demanding a
minimum sampling and a single bump. The remaining \pacz -like variable stars
show a strong correlation between the timescale and the flux
increase. In particular, bright ($\Delta R<21$) variations have timescales
between $30\,$days and $70\,$days, whereas we expect more than $80\%$ of the microlensing events to be
shorter than $25\,$days (for \macho s lighter than few solar
masses). So, for the moment, we focus on this best signal-to-noise ratio region:
$\Delta R<21$ and $t_{1/2}<25\,$days. This leads to 7 microlensing
candidates: \mbox{PA$\,$99-N1} \cite{auriere01,paulin03}, \mbox{PA$\,$99-N2} \cite{paulin03,an03} (the most
spectacular candidate, see \mbox{figure \ref{fig:n2-evts7-constraints}}),
\mbox{PA$\,$00-S3 \cite{paulin03}}, \mbox{PA$\,$00-S4 \cite{paulin02,paulin03}}, \mbox{PA$\,$00-S5 \cite{calchi04}}, \mbox{PA$\,$00-N6 \cite{calchi04}} and \mbox{PA$\,$99-S7 \cite{calchi04}}. Some of these lightcurves
are shown on figure \ref{fig:lightcurves}. Most of these
candidates were confirmed by other collaborations
\cite{riffeser03,dejong04}. We are working on additional statistical
criteria to search for fainter and/or longer events.

\section{No evidence for machos}
The detected microlensing signal is compatible with
self-lensing\footnote{As well as the \macho -signal, we expect a self-lensing
signal (when the lens is a star), clustered towards the M31 bulge.},
so there is no evidence for \macho s. The spatial distribution of the 7 microlensing
candidates is shown on the right panel of figure
\ref{fig:n2-evts7-constraints}. There are 4 events towards the bulge (\mbox{PA$\,$99-N1}, \mbox{PA$\,$00-S3}, \mbox{PA$\,$00-N6} and \mbox{PA$\,$99-S7}), one event
(\mbox{PA$\,$00-S4}) towards M32 (a dwarf galaxy orbiting around M31) and 2 events (\mbox{PA$\,$99-N2}
and \mbox{PA$\,$00-S5}) towards the disk. To compute constraints on
the content of M31 and the Milky Way halos, we consider the microlensing signal at more than $8'$ from the M31
center and more than $2'$ from M32. According to our preliminary
estimation of the detection efficiency, we expect 1 self-lensing event
in this area, and more than 20 (8) \macho{} events for standard halos full of \macho s
with masses between $10^{-4}\,M_\odot$ and $10^{-1}\,M_\odot$
($10^{-1}\,M_\odot$ and $1\,M_\odot$). The existence of 2 possible
\macho{} events
(\mbox{PA$\,$99-N2} and \mbox{PA$\,$00-S5}) leads to the following
upper limit: less than 25\% (60\%) of standard halos can be composed of
objects with masses between $10^{-4}\,M_\odot$ and $10^{-1}\,M_\odot$ ($10^{-1}\,M_\odot$ and
$1\,M_\odot$) at the 95\% C.L. This first constraint on \macho s in a distant spiral galaxy is currently improved (by lowering the
detection thresholds, improving the detection
efficiency, and including the fourth year data) and will be submitted
soon in Calchi Novati S., Paulin-Henriksson S. \mbox{et al. \cite{calchi04}}.

\end{document}